\documentclass[floatfix, prd, twocolumn, showpacs, showkeys, preprintnumbers, superscriptaddress,nofootinbib, 10pt]{revtex4-1}
\usepackage[usenames,svgnames,hyperref]{xcolor}
\usepackage{hyperref}
\hypersetup{colorlinks,linkcolor=blue,citecolor=red, urlcolor=blue}

\usepackage{graphicx}
\usepackage[sort&compress]{natbib}
\usepackage{subfigure}
\usepackage{amsmath}
\usepackage{amssymb}
\usepackage{amsfonts}
\usepackage{rotating}
\usepackage{cancel}
\usepackage{lipsum}
\usepackage{mathtools}
\usepackage{color}
\usepackage{bbm}
\usepackage{dsfont}
\usepackage{bbold}
\usepackage{multirow}
\usepackage[T1]{fontenc}
\usepackage{newtxtext, newtxmath}
\usepackage{pifont}
\usepackage{physics}
\usepackage{CJKutf8}
\usepackage{import}
\usepackage{tikz}
\usepackage{tikz-cd}
\usepackage{pgfplots}
\usepackage{comment}
\usepackage[normalem]{ulem}
\usepackage{verbatim}

\begin{document}

\title{Pole trajectories of the $\Lambda(1405)$ help establish its dynamical nature}
\author{Zejian~Zhuang~(\begin{CJK*}{UTF8}{gbsn}庄泽坚\end{CJK*})}
\email[Corresponding author: ]{zejian.zhuang@ific.uv.es}
\affiliation{Departamento de F\'{\i}sica Te\'orica and IFIC,
Centro Mixto Universidad de Valencia-CSIC, Parc Científic UV, C/ Catedrático José Beltrán, 2, 46980 Paterna, Spain} 
\author{Raquel~Molina}
\email[Corresponding author: ]{raquel.molina@ific.uv.es}
\affiliation{Departamento de F\'{\i}sica Te\'orica and IFIC,
Centro Mixto Universidad de Valencia-CSIC, Parc Científic UV, C/ Catedrático José Beltrán, 2, 46980 Paterna, Spain}   
\author{Jun-Xu Lu}
\email{ljxwohool@buaa.edu.cn}
\affiliation{School of Physics, Beihang University, Beijing 102206, China}
\author{Li-Sheng Geng}
\email{lisheng.geng@buaa.edu.cn}
\affiliation{School of Physics, Beihang University, Beijing 102206, China}
\affiliation{Beijing Key Laboratory of Advanced Nuclear Materials and Physics, Beihang University, Beijing 102206, China}
\affiliation{Peng Huanwu Collaborative Center for Research and Education, Beihang University, Beijing 100191, China}
\affiliation{Southern Center for Nuclear-Science Theory (SCNT), Institute of Modern Physics, Chinese Academy of Sciences, Huizhou 516000, China}

\begin{abstract}  
The $\Lambda(1405)$ has been one of the most controversial exotic baryons. If the $\Lambda(1405)$ possesses a two-pole molecular structure, these poles are expected to evolve differently towards the SU(3) limit.  
   From an analysis of a recent LQCD simulation on the $\pi\Sigma-\bar{K}N$ scattering for $I=0$ and the study of the quark mass dependence of the octet baryon masses, we determine for the first time the trajectories of these poles towards the symmetric point over the $\mathrm{Tr}[M]=C$ trajectory accurately. At $m_\pi\simeq 200$ MeV, our results are consistent with the lattice simulations, and the extrapolations to the physical point, based on the NLO chiral Lagrangians, agree well with existing experimental analyses. We predict qualitatively similar trajectories at LO and up to NLO, consistent with the LO interaction's dominance. At the SU(3) symmetric point of this trajectory, both poles are on the physical sheet, and the lower pole is located at $E^{(1)}=1573(6)(6)$ MeV, becoming a SU(3) singlet, while the higher pole at $E^{(8a)}=1589(7)(5)$ MeV couples to the octet representation. Moreover, we make predictions in $I=1$ for the $\Sigma^*$ resonance. We find a resonance pole that evolves into a bound state around $m_\pi=415$~MeV in this sector. The results presented here are crucial to shed light on the molecular nature of exotic strange baryon resonances and can be tested in future LQCD simulations.
\end{abstract}
\maketitle
\section{Introduction}
The $\Lambda(1405)$, discovered in bubble chamber experiments at low-energy $K^-p$ scattering as a resonance decaying into $\pi^-\Sigma^+$~\cite{Dalitz1,Dalitz2}, has been a subject of controversy for a long time. This can be mainly attributed to the two-pole structure predicted in the chiral unitary approach, which refers to a lower pole above the $\pi\Sigma$ threshold and a higher pole below the $\bar{K}N$ threshold~\cite{Oset1997,Oller2000,Jido2003,Borasoy2005,Borasoy2006,Hyodo2007,Ikeda2011,Ikeda2012,Guo2012,Feijoo2015,Feijoo2018,Sadasivan2018,Molina2023}. Only recently, the lower pole associated to the two-pole structure demanded by chiral dynamics and SU(3) flavor symmetry has been included in the PDG~\cite{pdg} about 60 years later.

The $\Lambda(1405)$ is also one of the first exotic resonances discovered. Note that in the recent years, there has been a proliferation of such exotic particles, like the $X(3872)$, the $XYZ$ states, the tetraquark $T_{cc}(3875)^+$ or the pentaquark states $P_c$ observed by experimental facilities like Belle, Babar or the LHCb~\cite{Belle:2003nnu,LHCb:2015yax, LHCb:2021vvq}. See also Ref.~\cite{Guo:2017jvc} for a review on hadronic molecular candidates. These major discoveries challenge our understanding of the hadron spectrum, reflecting the complexity allowed by QCD, the theory of the strong interactions, which represents the key for the formation of matter. In particular, the study of the $\Lambda(1405)$ can also help to understand the interaction of strange particles with nuclear matter, which is relevant for studying the production of strange hadrons in neutron stars and also to learn about the strangeness content of the proton~\cite{Gasser1990}. 

The combination of the chiral lagranginas with unitarity constraints has a long history of success. On the one hand, the chiral lagrangians are based on an expansion of the light degree's of freedom, the pion mass and momentum, satisfying chiral symmetry, the symmetry of the QCD lagrangian at zero quark masses~\cite{Gasser:1983yg,Gasser1984,Weinberg:1990rz,Weinberg:1991um,Pich1995,Ecker1994,Bernard1995,Meissner1993}. On the other hand, unitarity is a requirement on the imaginary part of the scattering amplitude from Scattering Theory~\cite{Martin:1970hmp,Friedrich:2015obs}. The combination of both has allowed to extend the use of chiral lagrangians to describe cross sections in the intermediate energy region. Some well-known examples of resonances that can be described in the chiral unitary approach are the scalar resonances, such as the $\sigma$ and the $\kappa$~\cite{Truong:1988zp,Dobado:1989qm,Oller:1998hw,Nieves:1998hp,GomezNicola:2001as}. In the baryon sector, the $N^*(1535)$ and the $\Lambda(1405)$ can also be well described within this approach~\cite{Oset1997,Jido2003,Kaiser1995,Kaiser1996,Nieves2001}. 

The $\Lambda(1405)$ quantum numbers are $J^P=\frac{1}{2}^-$~\cite{Clas2014}. Surprisingly, being the first negative parity excitation of the $\Lambda$, it is lighter than its nucleon counterpart by around $100$~MeV. This can be naturally explained in the chiral unitary approach, considering the $N^*(1535)$ and the $\Lambda(1405)$ as dynamically generated resonances from the $K\Lambda$, $K\Sigma$ channels, and the $\bar{K}N$, $\pi\Sigma$ channels respectively, see Refs.~\cite{Oset1997,Jido2003,Kaiser1995,Kaiser1996,Nieves2001,Molina2023}. The interaction in these approaches is driven by the lowest order (LO) interaction in $s$-wave. This Weinberg-Tomozawa term is attractive and leads to two poles in the unphysical sheet. At NLO the two-pole structure also persists~\cite{Jido2002,Cieply2015,Feijoo2018,Sadasivan2018,Mai2014,Mai2012,Cieply2016}. Some of the NLO low-energy constants (LECs) in the chiral lagrangian are constrained by the splitting of the baryon masses, the pion-nucleon sigma term, and the strangeness content of the proton~\cite{Gasser1990}. Yet, because of the unitarization process, these are usually kept as free parameters adjusted to experimental data. On the other hand, there are experimental data from DEAR, LEPS, CLAS, COSY, HADES~\cite{DEAR:2005fdl,Niiyama2008,Clas2013,Clas2014,Clas2013b,Zychor2007,Hades2012}, which have been used to fix these NLO parameters in theoretical analyses~\cite{Oller:2005ig,Hassanvand2012,Roca2013,Bayar2017}. As a result of these studies, the position of the higher pole is well constrained, while the lower one bears large uncertainties. Recent NNLO studies have further supported the two-pole structure~\cite{Lu2022}. Both pole positions are determined more precisely by including the experimental data on $\bar{K}N$, $KN$, and $\pi N$ cross sections. This analysis yields,  $E_1=[1392\pm 8-i\, (102\pm 15)]$~MeV, and, $E_2=[1425\pm 1-i\,(13\pm 4)]$ MeV. In all the studies based on the chiral unitary approach, the first pole couples more strongly to $\pi\Sigma$, while the second one, also seen in $\pi\Sigma$, does more to $\bar{K}N$. In Ref.~\cite{Liu:2016wxq}, the analysis of experimental cross sections with Hamiltonian effective field theory leads also to the two-pole structure regardless of a possible bare component, and with the finite-volume spectrum consistent with LQCD. The light-quark contribution to the $\Lambda(1405)$ magnetic form factor from LQCD also support its molecular nature~\cite{Hall:2016kou}. The $K^- p$ interaction and the presence of the two-pole structure have also been inferred from the measured femtoscopic correlation functions~\cite{Alice2019,Molina2023,Alice2022}. However, there is still controversy since the recent J-PARC line-shape data seem only to require the presence of one resonance~\cite{Jparc2022}. For recent reviews of the $\Lambda(1405)$, see Refs.~\cite{reviewmai,reviewmeissner}. 

In the past, the scarce lattice QCD (LQCD) simulations investigating the $\Lambda(1405)$ have only considered single baryon three quark interpolating fields~\cite{Gubler2016,Menadue2011,Engel2012,Engel2013,Nemoto2003,Burch2006,Takahashi2009,Meinel2021,Hall2014}. The difficulties in extracting the two poles from the LQCD data were emphasized in Ref.~\cite{MartinezTorres:2012yi}. However, a recent LQCD simulation included the meson-baryon operators. These results were most welcome since, for the first time, a signature of the lower pole of the $\Lambda(1405)$ has been observed as a virtual bound state at a pion mass of $\sim 200$~MeV~\cite{Bulava2023,Bulava2023b}. The pole positions obtained in this simulation~\cite{Bulava2023,Bulava2023b} are, $E_1=1392(9)(2)(16)$~MeV, and $E_2=1455(13)(2)(17)-i\,11.5(4.4)(4)(0.1)$~MeV. The lower pole couples more to $\pi\Sigma$ with approximately double strength than to $\bar{K}N$, and the higher to $\bar{K}N$, with ratios of the couplings in line with earlier results based on the chiral unitary approach. 

Nonetheless, in Refs.~\cite{Bulava2023,Bulava2023b}, the parameterization used to extract the poles is based on the effective range expansion (ERE) and the $K$-matrix formalism, neglecting the real part of the loop function~\cite{Oset1997,Jido2003}. In addition, a chiral extrapolation to the physical point from the analysis of the energy levels of Refs.~\cite{Bulava2023,Bulava2023b} to investigate the compatibility of the result with the experimental data has not yet been performed. The chiral unitary approach predicts different trends of the two poles of the $\Lambda(1405)$ towards the SU(3) limit. In this limit, while the low-energy pole belongs to the singlet representation, the higher one does to the octet~\cite{Jido2003,Garcia-Recio:2003ejq,Bruns:2021krp}. Here, we predict the trends of these poles towards the SU(3) limit from the LQCD data analysis of Refs.~\cite{Bulava2023,Bulava2023b} within the chiral unitary approach. The chiral path studied here is of the type $\mathrm{Tr}[M]=C$~\cite{Bulava2023,Bulava2023b}.\footnote{Here, $M$ stands for the quark mass matrix, and the trajectory $\mathrm{Tr}[M]=C$ means $m_u+m_d+m_s=C$.} These trajectories are a fundamental property of the $\Lambda(1405)$ connected to the SU(3) breaking pattern and its nature as a dynamically generated resonance.

The first explorations of the trajectories of the $\Lambda(1405)$ with increasing pion masses can be found in Refs.~\cite{Molina2015,Pavao2020,Guo2023,Xie2023,Ren2024}. In Refs.~\cite{Molina2015,Pavao2020}, earlier LQCD simulations were compared with the predictions of the chiral unitary approach, concluding that it was important to include meson-baryon interpolating fields in future simulations. The pole trends from the SU(3) limit to the physical values are investigated in Ref.~\cite{Guo2023} using the chiral unitary approach at NLO with LECs fixed to the experimental data on $K^-p$ scattering~\cite{reviewmai}. 
These trajectories are also investigated in Refs.~\cite{Xie2023,Ren2024}. 

The simulations of Refs.~\cite{Bulava2023,Bulava2023b} are based on CLS ensembles. The light-baryon spectrum is extracted from LQCD for these ensembles in Ref.~\cite{RQCD2022}. To conduct the extrapolation to the physical point, the authors of Ref.~\cite{RQCD2022} employ Baryon Chiral Perturbation Theory (BChPT). See also Ref.~\cite{Young:2002cj} for extrapolations of the baryon masses within quenched ChPT.

 In any of the previous works~\cite{Guo2023,Xie2023,Ren2024}, neither the energy levels for the $\bar{K}N$ and $\pi\Sigma$ scattering~\cite{Bulava2023,Bulava2023b} nor the most recent LQCD data of baryon masses~\cite{RQCD2022} for the CLS ensembles have been analyzed. This step is essential to understand the SU(3) symmetry pattern of the $\Lambda(1405)$ two poles, their nature, and the LQCD and experimental data consistency. In this work, we fill this gap by conducting the first extrapolation to the physical point from the analysis of the energy levels of Refs.~\cite{Bulava2023,Bulava2023b} and study the trajectories of the two poles towards the SU(3) symmetric line along the $\mathrm{Tr}[M]=C$ curve. This is a crucial test for the chiral unitary approach in future LQCD simulations and the two-pole structure of the $\Lambda(1405)$. Our work is based on the one-loop NLO covariant baryon chiral perturbation theory. We will also study the effect of the NLO terms on these trajectories. Predictions for both kind of trends of these poles, $\mathrm{Tr}[M]=C$ and  $m_s=m_{s,\mathrm{phy}}$, are made here. We show that these are indeed very different due to the dynamical nature of the $\Lambda(1405)$.


\section{Formalism}~\label{sec:for}
The SU(3) chiral Lagrangians for the meson-baryon interaction are given in Refs.~\cite{Pich1995,Ecker1994,Bernard1995,Meissner1993}. For \textit{isospin, strangeness}, $I=0, S=-1$, we consider the four channels $i,j=\pi\Sigma, \bar{K}N, \eta\Lambda, K\Xi$. The interaction kernel up to NLO reads,
\begin{equation}\label{eq:pot}
    V_{ij}=V^{\mathrm{LO}}_{ij} + V^\mathrm{NLO}_{ij},
\end{equation}
where $V^\mathrm{LO}=V^\mathrm{WT} + V^\mathrm{Born}$. The $V^\mathrm{WT}$ and NLO term are given by~\cite{Feijoo2015},
\begin{align}
    V_{ij}^\mathrm{WT} &= -\frac{N_i N_j}{4f^2}\left[C_{ij}\qty(2\sqrt{s}-M_i-M_j)\right],\label{eq:wt-term} \\
    V_{ij}^{\mathrm{NLO}} &= -\frac{N_i N_j}{f^2}\qty(D_{ij}-2k_\mu k'^\mu L_{ij}),\label{eq:nlo-term}
\end{align}
where $N_i=\sqrt{(M_i+E_i)/2M_i}$, and $M_i,E_i$ stand for the baryon mass and energy of channel $i$, respectively. Eq.~(\ref{eq:wt-term}) represents the Weinberg-Tomowawa (WT) interaction (LO), while the Eq.~(\ref{eq:nlo-term}) comes from the NLO lagrangian. We include the Born term in Eq.~(\ref{eq:pot}) in the $s$- and $u$-channels for the $J^P=1/2^+$ octet baryon-exchange diagrams, which are taken from Ref.~\cite{PhysRevC.100.015208}. The $u$-channel Born term can lead to a contribution of similar size than the NLO term for some of the channels~~\cite{PhysRevC.100.015208}. In Eq.~(\ref{eq:nlo-term}), the $D_{ij}$ and $ L_{ij}$ coefficients are a combination of the pertinent LECs of $b_0, b_D, b_F$, and $d_i$, $i=1,4$.  The LECs $b_i$ and $d_i$ stem from the NLO contact terms of the chiral lagrangians for the meson-baryon interaction~\cite{Frink2006,Oller2006}. The LECs $b_i$ are related to the splitting of the octet baryon masses and pion-nucleon sigma term~\cite{Gasser1990} since they come from the Lagrangian terms proportional to the quark mass, and the LECs $d_i$ appear only in the $P_iB_j\to P_kB_l$ scattering ($P=$ pseudoscalar meson, $B=$ baryon). They contribute to the non-relativistic limit of the $s$-wave interaction. These combinations are given in Appendix A of Ref.~\cite{Feijoo2015}. See also Ref.~\cite{Borasoy2005}. We consider the one-loop covariant baryon chiral perturbation theory for the latter. The LECs $b_i$ are fitted to the recent RQCD simulations on the quark mass dependence of baryon masses for the CLS ensembles~\cite{RQCD2022}, which include data on the $m_s=m_{s,\mathrm{phy}}$, $\mathrm{Tr}[M]=C$ and $m_{u(d)}=m_s$ trajectories. See Fig.~\ref{fig:baryon-mass1}. The fit in this work is discussed in more detail in the Supplemental Material. Table~\ref{tab:LECs-bi} gives the values obtained for these LECs-$b_i$. The baryon mass obtained in the chiral limit is $m_0 = 805(40)(40)$ MeV. In Fig.~\ref{fig:baryon-mass1}, the symmetric line $m_s=m_l$, with $l=u,d$, is shown in green color. For the case of the type of trajectory studied here, $\mathrm{Tr}[M]=C$, the SU(3) limit is reached at $m_\pi=423$~MeV with a baryon mass $m_B= 1182(7)(5)$~MeV. \footnote{Note that this point is different for the case of the $m_s=m_{s,\mathrm{phy}}$ trajectory, which is around $m_\pi=760$~MeV~\cite{Ren2012}. }

\begin{figure}[!htbp]
    \includegraphics[width=0.8\columnwidth]{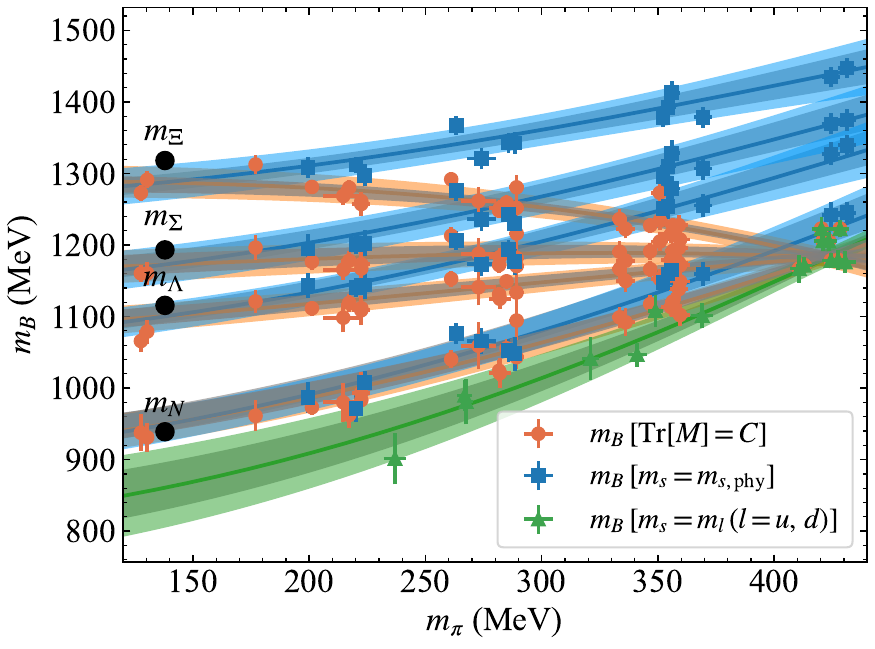}
\caption{\label{fig:baryon-mass1}Fits to the hadron masses. The data is provided by RQCD~\cite{RQCD2022}. The circles, squares, and triangles denote the $\mathrm{Tr}\qty[M]=C$, $m_s=m_{s,\mathrm{phy}}$ trajectories, and the symmetric line $m_s=m_l$, where $l=u,d$, respectively. The circles in black are the physical masses of the hadrons. The colored error bands consider the systematic and statistical errors of the hadron masses. The error bands in gray take into account only statistical errors.}
\end{figure}

On the other hand, the masses of the pseudoscalar mesons are calculated up to NLO in ChPT. The quark mass dependence of the pseudoscalar meson masses $m_\phi$ and decay constants $f_\phi$, with $\phi=\pi,K,\eta$, were investigated in~\cite{Molina2020} by analyzing the recent LQCD data including the CLS ensembles. In the present work, we use the NLO ChPT pseudoscalar masses and take the values of the LECs $L$'s obtained from the global fit,  see Table X of ~\cite{Molina2020}. Since both light and strange mesons are present in the channels $i$, $i=1,4$, we take the average $f=\qty(f_\pi+f_K+f_\eta)/3$ in Eq.~\eqref{eq:pot}. Therefore, the pion mass dependence on the decay constants is incorporated in our framework, taking into account recent LQCD data.
\begin{table}[!htpb]
    \renewcommand{\arraystretch}{1.8}
\begin{tabular}{lccc}
    \hline
    ~ & $b_0$ & $b_D$ & $b_F$ \\
    \hline
    This work & $-0.665(40)(28)$ & $0.062(26)(8)$ & $-0.354(18)(9)$ \\
    BChPT FV~\cite{RQCD2022} & $-0.739_{(84)}^{(70)}$ & $0.056_{(39)}^{(43)}$ & $-0.44_{(26)}^{(40)}$ \\
    \hline
\end{tabular}
\caption{\label{tab:LECs-bi}Values of the LECs-$b_i$ in units of $\mathrm{GeV}^{-1}$. The first uncertainty is statistical and the second one is systematic from the lattice spacing~\cite{RQCD2022}.}
\end{table}

The scattering amplitude in the infinite volume can be written as 
\begin{equation}
 T^{-1}=V^{-1}_{0} -G\ ,\label{eq:bethe}
\end{equation}
where $V_{0}\equiv V_{(L=0)}$ is the $s$-wave projection of Eq.~(\ref{eq:pot}), and $G$ is a diagonal matrix with the meson-baryon loop functions, whose elements are
\begin{eqnarray}
 G_j(P)=2iM_j \int\frac{\dd^4q}{(2\pi)^4}\frac{1}{q^2-m^2_j+i\epsilon}\frac{1}{(P-q)^2-M_j^2+i\epsilon}\ .\nonumber\\\label{eq:loop}
\end{eqnarray}
where $P$ stands for $P^\mu$, with $s=P^2$, the squared energy in the center-of-mass (CM) frame. The above equation can be evaluated using the dimensional regularization (DR) method or a cutoff $q_{\mathrm{max}}$~\cite{Oller1998}. Comparing the expressions obtained in the two schemes, one can obtain the dependence of the subtraction constant on the masses of the mesons and baryons in the loop. Here, we follow the relation obtained in Ref.~\cite{Oller:2019opk}, where, 
\begin{equation}\label{eq:subtraction-cutoff}
    \begin{aligned}
        a(\mu) &= -\frac{2}{m_j+M_j}\left[m_j\log\qty(1 + \sqrt{1 + \frac{m_j^2}{{q}_\mathrm{max}^2}})\right.\\
        &\left.+M_j\log\qty(1 + \sqrt{1 + \frac{M^2_j}{{q}_\mathrm{max}^2}})\right] + 2\log\qty(\frac{\mu}{q_\mathrm{max}}).
    \end{aligned}
\end{equation}
and we set the scale $\mu=630$ MeV~\cite{Oset1997}.\footnote{The parameter $\mu$ in Eq.~(\ref{eq:subtraction-cutoff}) is a free parameter that should be of the order of the interaction range, that is $1$~GeV.  This parameter is related to $\alpha$, the subtraction constant. With the choice of $\mu\simeq 630$~MeV, one obtains a subtraction constant of the order of $-2$, which is the natural size for molecular states~\cite{Oller2000}.}

On the other hand, in the finite volume, neglecting the effects from higher partial waves, the scattering amplitude reads,
\begin{equation}
 \tilde{T}^{-1}=V^{-1}_{0} -\tilde{G}\ .\label{eq:bethefin}
\end{equation}
The finite-volume loop function $\tilde{G}$ can be calculated in the DR scheme as~\cite{MartinezTorres2011}, \begin{eqnarray}\label{eq:gt}
 \widetilde{G}(P)=G^\mathrm{DR}(P)+\lim_{{q}'_\mathrm{max}\to\infty}\Delta G(P,{q}'_\mathrm{max}),
\end{eqnarray}
being $\Delta G = \tilde{G}^\mathrm{co} - G^\mathrm{co}$, and the loop function in the infinite volume, $G^\mathrm{DR}$, $G^\mathrm{co}$, are evaluated in the dimensional regularization (DR) and cutoff scheme (co), respectively~\cite{Molina2015}. In moving frames, $\tilde{G}^\mathrm{co}$ reads~\cite{Doring2012}\footnote{Note that Lorentz symmetry is broken in the finite volume.},
\begin{eqnarray}
 \tilde{G}^\mathrm{co}(P,{q}_\mathrm{max})=\frac{2M}{L^3}\sum_{\vec{n}}^{q_\mathrm{max}}\frac{E}{P_0}I(|\vec{q}\,^*(\vec{q})|)\ ,\label{eq:g00}
\end{eqnarray}
 In the above equation, $q^*$ is the CM momentum, $\vec{q}_1\,^*+\vec{q}_2\,^*=0$, which can be written in terms of the momenta in the boosted frame, $\vec{q}=2\pi\frac{\vec{n}}{L}$, with $P^\mu=q_1^\mu+q_2^\mu$, the four-momenta of the system. The function $I(q)$ reads,
\begin{eqnarray}
 I(\vec{q}\,)=\frac{\omega_1(q)+\omega_2(q)}{2\omega_1(q)\omega_2(q)\left[P^2_0-(\omega_1(q)+\omega_2(q))^2+i\epsilon\right]}\ ,
\end{eqnarray}
with $\omega_i=\sqrt{q^{2}+m_i^2}$, $i=1,2$ refer to the meson ($m_1=m$) and baryon ($m_2=M$) in the loop, and $q=|\vec{q}\,|$. Neglecting the effects from higher partial waves, the energy levels are given solely by~\cite{Doring2012},
\begin{equation}
 \det\qty[I-V_0\tilde{G}]=0.\label{eq:ener}
\end{equation} 
Near the resonance region around $\sqrt{s_0}$, the coupling $g_i$ to the $i$th meson-baryon channel can be easily evaluated since the amplitude behaves as
\begin{equation}\label{eq:t_coupling}
    T_{ij} \simeq \frac{g_i g_j}{\sqrt{s}-\sqrt{s_0}}.
\end{equation}
Once the interaction $V_0$ is obtained from a fit to the energy levels through Eq.~(\ref{eq:ener}), phase shifts can be readily obtained through Eq.~(\ref{eq:bethe}),
\begin{eqnarray}
 p\mathrm{cot}\delta_j=-\frac{8\pi E}{2 M_j}(T_{jj})^{-1}+i\,p_j\ .\label{eq:phase}
\end{eqnarray}
Note that, in the one channel case, the phase shifts at the LQCD energies obtained by solving Eq.~(\ref{eq:bethefin}) depend only on the second term of Eq.~(\ref{eq:gt}), which is parameter free, and thus, these turn out to be model independent, similarly as in the L\"uscher approach~\cite{Luscher:1986pf,Luscher:1990ux}.  
\section{Results and discussions\label{sec:result}}
\begin{figure}[!htbp]
    \includegraphics[width=0.85\columnwidth]{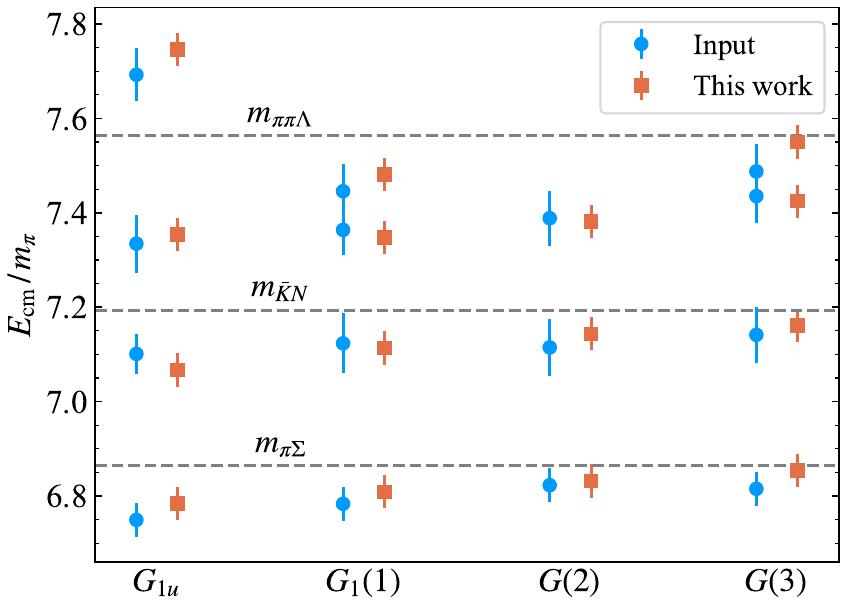}
\caption{\label{fig:finite-volume-spectrum}Finite-volume spectrum~\cite{Bulava2023,Bulava2023b} in the center-of-mass frame used as inputs to constrain the LECs of the chiral Lagrangian up to NLO.}
\end{figure}

We analyze the energy levels of Refs.~\cite{Bulava2023,Bulava2023b} for $m_\pi\simeq 200$ MeV through Eq.~\eqref{eq:ener}, with the $s$-wave projection of the NLO interaction in Eq.~\eqref{eq:pot}. The $\chi^2$ is given by 
\begin{equation}\label{eq:chi-square}
    \chi^2=\Delta E^T C^{-1}_E \Delta E,
\end{equation}
where $C_E$ is the covariance matrix for the energy levels, which is evaluated from the raw data of Ref.~\cite{Bulava2023}, and $\Delta E_i=E_i' - E_i$ is the difference between the lattice energy level $E_i$ and the predicted one, $E_i'$. The resulting energy levels are shown in Fig.~\ref{fig:finite-volume-spectrum}. The LQCD energy levels are very well reproduced by the interaction up to NLO. We have included the first four energy levels in the fit, with all data points shown in Fig.~\ref{fig:finite-volume-spectrum}.  The LECs-$b_i$ are determined by reproducing the quark mass dependence of the baryon masses along the trajectories of the CLS ensembles~\cite{RQCD2022}, and the $d_i$ are fitted to the energy levels. The value of the reduced-$\chi^2$ obtained is $\chi_{\mathrm{dof}}^2=2.3$. ~\footnote{We do not consider systematic errors in the LQCD data when performing the fit.} If the correlation matrix between the energy levels is omitted, the value obtained is $\chi_{\mathrm{dof}}^2=0.7$. Thus, the correlation matrix significantly constrains the NLO LECs. The results of the LECs $d_i$ are given in Table~\ref{tab:LECs-di-cutoff}. The cutoff $q_\mathrm{max}$ in Eq.~\eqref{eq:subtraction-cutoff} obtained in this fit is $623(23)(23)$ MeV.
\begin{table}[!tbph]
\renewcommand{\arraystretch}{1.8}
\setlength{\tabcolsep}{0.2cm}
    \centering
    \begin{tabular}{ccccl}
    \hline
    $d_1$ & $d_2$ & $d_3$ & $d_4$\\
    \hline
    $-0.38(9)(7)$ & $0.02(1)(1)$ & $-0.07(3)(3)$ & $-0.45(5)(6)$\\
    \hline
    \end{tabular}
    \caption{\label{tab:LECs-di-cutoff}Results for the LECs $d_i$ in units of $\mathrm{GeV}^{-1}$. The second error comes from the uncertainty in the lattice spacing~\cite{Bulava2023,Bulava2023b}.}
\end{table}

In the infinite volume limit, we find two poles related to the $\Lambda(1405)$, one is a virtual pole and the other a resonance.  In Table~\ref{tab:strategy1-up-to-NLO-fitting-result-st1}, we give the positions and couplings of the poles to the different channels for $m_\pi=138$, $200$ MeV, in the first and second rows for the four-channel coupling calculation. When the $\eta\Lambda$ and $K\Xi$ channels are removed, the virtual resonance related to the lower pole becomes a virtual bound state. The pole positions and couplings for the two-coupled channel system are shown in the third row, which can be compared with the LQCD simulation results~\cite{Bulava2023,Bulava2023b}. The positions of the two poles at $m_\pi\simeq 200$ MeV agree very well with the values obtained in LQCD for the two-coupled channel analysis, which are $z_1=1392(9)(2)(16)$ MeV and $z_2=1455(13)(2)(17)-i11.5(4.4)(4)(0.1)$. The ratios of the couplings obtained in this work (third row) have an excellent agreement with the ones from LQCD ~\cite{Bulava2023,Bulava2023b}, 
\begin{equation}\label{eq:radio_coupling}
    \qty|\frac{g_{\pi\Sigma}^{(1)}}{g_{\bar{K}N}^{(1)}}| = 1.9(4)_\mathrm{st}(6)_\mathrm{md},\ 
    \qty|\frac{g_{\pi\Sigma}^{(2)}}{g_{\bar{K}N}^{(2)}}| = 0.53(9)_\mathrm{st}(10)_\mathrm{md},
\end{equation}
While the first pole couples strongly to $\pi\Sigma$, the second one does to $\bar{K}N$. Overall, the four-coupled-channel calculation agrees well with the two-coupled channel results, but the imaginary part of the lower pole can vary significantly, indicating that it can become a virtual resonance. The couplings to the $\eta\Lambda$ and $K\Xi$ 
 are also non-negligible. Note that these channels have been neglected in the two-coupled-channel LQCD analysis~\cite{Bulava2023,Bulava2023b}. 

\begin{figure}[!htbp]
	\includegraphics[width=\columnwidth]{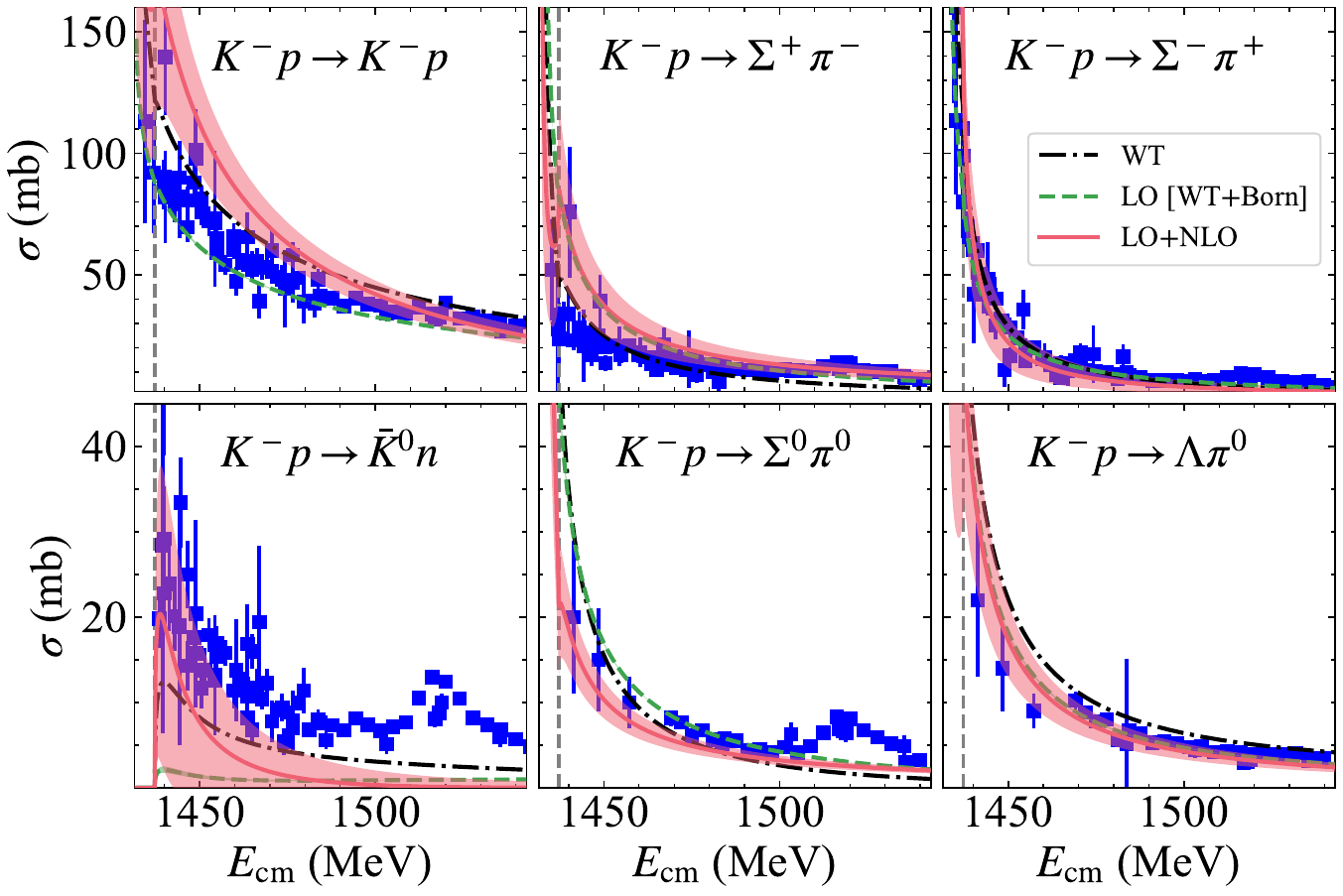}
	\caption{\label{fig:exp-data}Comparison of the predicted cross-sections and the experimental data~\cite{reviewmai}. The error bands of the cross sections are constrained by the correlation matrix of the fit to the LQCD energy levels.}
\end{figure}

Within our approach, we can extrapolate the results to the physical point. In Fig.~\ref{fig:pole_position-conparsion}, we show the extrapolation to the physical point of the two poles compared to previous experimental data analysis. These agree well with experiment. In particular, the position of the higher pole is well constrained. Furthermore, we can compare our results with the experimental cross sections available in Ref.~\cite{reviewmai}. Remarkably, without conducting a fit to these data, our results agree well with the experimental cross sections. See Fig.~\ref{fig:exp-data}. Hence, agreement between the LQCD, experimental data, and the chiral unitary approach is found for the first time. This strongly supports the quark-mass dependence obtained here, which is determined reliably and includes all sources of quark mass dependence. Notice that in previous works, only qualitative results were obtained~\cite{Jido2003, Guo2023,Xie2023,Ren2024,Sadasivan:2022srs}. For $m_\pi\simeq  200$ MeV, we also obtain phase shifts that resemble the ones obtained in Refs.~\cite{Bulava2023,Bulava2023b} as shown in the Supplemental Material.

Next, we comment on the results obtained in the SU(3) limit. In this limit, we obtain a resonance coupling to the singlet representation corresponding to the lower pole and two poles coupling to the octet representations. Note that, in principle, the pseudoscalar meson-baryon channels couple to both, $8$ and $8'$ representations (see the Supplemental Material)\footnote{In particular, the $\bar{K}N$ channel couples to both octet representations, $8$ and $8'$.}, and therefore, the resonance generated by them can also couple to both, mixing these representations~\cite{Bruns:2021krp}. In the Supplemental material, we also provide the kernel of the interaction, for the WT, Born, and NLO, in the SU(3) representation basis. However, since the SU(3) symmetry is respected in this limit, it is possible to find a basis where the octet representations disentangle by diagonalizing the interaction\footnote{In the language of Quantum Mechanics, since the Hamiltonian conmutes with the generators of SU(3), $\lambda_i$, it is possible to find common eigenvectors of the Hamiltonian and the generator of the rotations in SU(3).}. Thus, we define
\begin{equation}
\begin{aligned}
    \ket{8a}&=\mathrm{cos}\,\theta\ket{8}_{I=0}-\mathrm{sin}\,\theta\ket{8'}_{I=0},\\
    \ket{8b}&=\mathrm{cos}\,\theta\ket{8}_{I=0}+\mathrm{sin}\,\theta\ket{8'}_{I=0},
\end{aligned}
\end{equation}
and to find the mixing angle we follow \cite{Bruns:2021krp}. We have obtained $\theta\simeq -\,60^\circ$. The pole positions and couplings obtained in this way are shown in Table~\ref{tab:pole-coupling-su3}. With only WT, we find one pole located at $E^{(1)}=1556(2)$ and two degenerated poles at $E^{(8a)}=E^{(8b)}=1606(1)$ MeV, coupling to the singlet and octet representations, respectively. Note that this degeneracy is \textit{accidental} caused by the WT interaction. When the Born and NLO terms are included, the two higher poles split in energy. This phenomena is in agreement with findings from previous works ~\cite{Jido2003,Bruns:2021krp,Guo2023}. In particular, while the lower pole couples to the $8a$, the higher one does to the $8b$ representation. The pole positions and couplings to the different representations are shown in Table~\ref{tab:pole-coupling-su3}. The mass shift between the two higher poles after including the NLO obtained in this work is $\Delta E^{(8)}=14$ MeV. 
\begin{figure}[!htbp]
    \resizebox{0.9\columnwidth}{!}{\import{}{./figure/pole_position_comparsion.pgf}}

\caption{\label{fig:pole_position-conparsion}The positions of the two $\Lambda(1405)$ poles (color in black) up to NLO obtained in the present study in comparison with previous works~\cite{Guo2012,Ikeda2012,Mai2014,Sadasivan2018,Cieply:2011nq,Shevchenko:2011ce,Haidenbauer:2010ch,Guo2023,Lu2022,Sadasivan:2022srs}.}
\end{figure}

\begin{figure*}[!htbp]
    \includegraphics[width=0.95\textwidth]{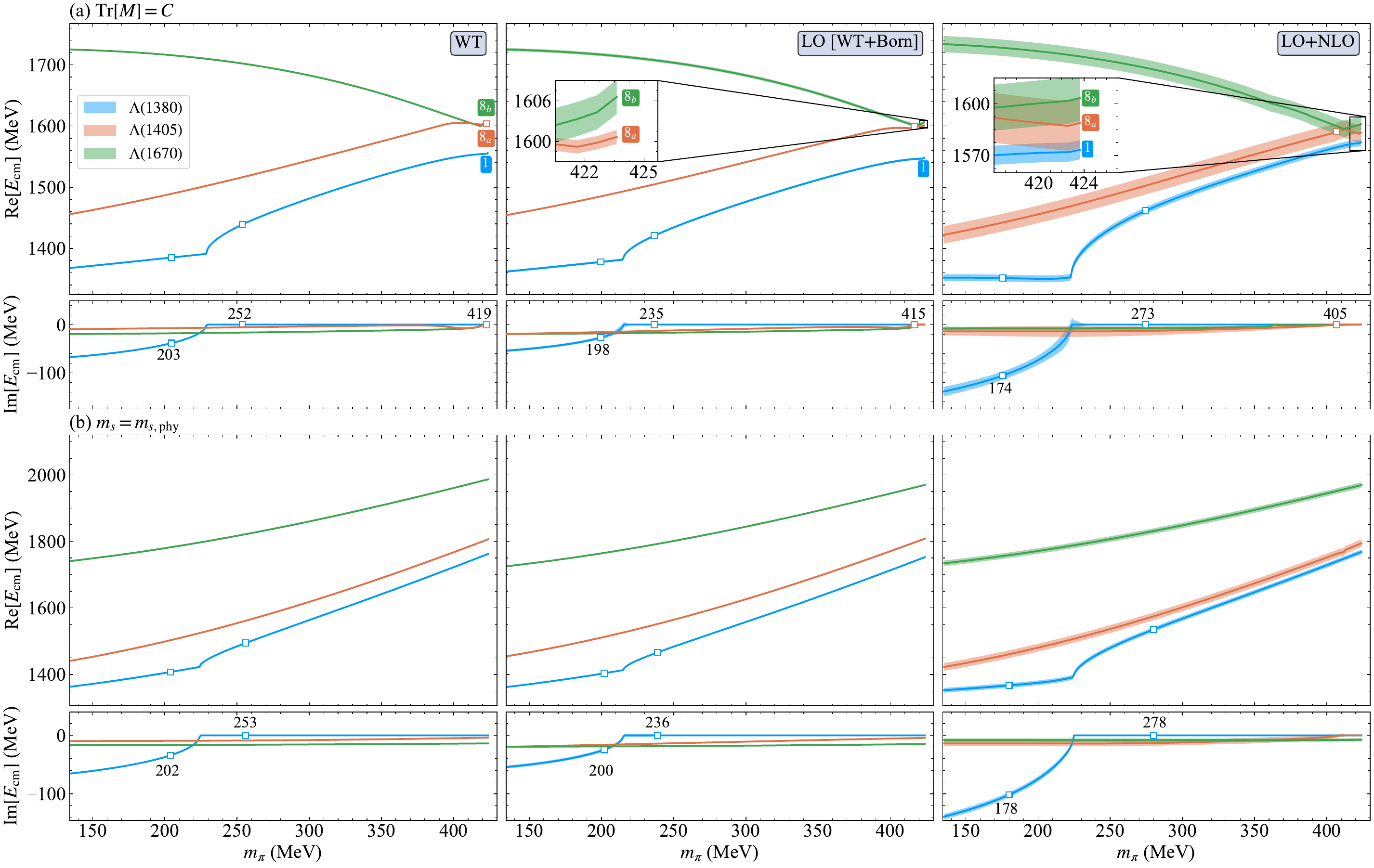}
\caption{\label{fig:trajectory}Trajectories of the three poles up to NLO for the $\Lambda(1405)$ and $\Lambda(1670)$ along the curve $\mathrm{Tr}[M]=C$~(a) and the curve $m_s=m_{s,\mathrm{phy}}$~(b). The uncertainties of trajectories up to NLO originate from the statistical errors and the lattice spacing error.}
\end{figure*}

Next, we show our prediction for the trajectories of the poles towards the SU(3) limit over the $\mathrm{Tr}[M]=C$ trajectory in Fig.~\ref{fig:trajectory}~(a), where we show the results at LO and up to NLO. We see that, at LO (LO+NLO), the lower pole evolves from a resonance into a virtual state at $m_\pi=198\ (174)$ MeV. Then, it becomes a bound state at $m_\pi=235\ (273)$ MeV and a singlet pole when it reaches the symmetric line. On the other hand, the higher pole changes from a resonance to a bound state at $m_\pi=415\ (405)$ MeV, close to the symmetric point. The higher pole of the $\Lambda(1405)$ and the $\Lambda(1670)$ couple to the $8a$ and $8b$ representations in the SU(3) limit. 
\begin{table*}[!tbph]
    \renewcommand{\arraystretch}{1.6}
 \setlength{\tabcolsep}{0.27cm}
    \centering
    
    \begin{tabular}{lccccccc}
    \hline 
    ~&$m_\pi$ (MeV) & Pole (MeV) & $\abs{g_{\pi\Sigma}}$ & $\abs{g_{\bar{K}N}}$ & $\abs{\frac{g_{\pi\Sigma}}{g_{\bar{K}N}}}$ & $\abs{g_{\eta\Lambda}}$ & $\abs{g_{K\Xi}}$\\\cline{2-8}
    \multirow{4}{*}{4 channels}&\multirow{2}{*}{$138$} & $1376(10)(10)-i142(19)(5)$ & $2.8(1)(2)$ & $2.3(6)(3)$ & $1.2(5)(1)$ & $1.5(3)(3)$ & $1.0(1)(2)$\\
                        ~ & ~ & $1418(11)(2) - i11(6)(3)$ & $1.1(5)(2)$ & $3.0(2)(2)$ & $0.4(1)(1)$ & $2.0(2)(5)$ & $0.4(1)(3)$ \\\cline{3-8}
                        ~ & \multirow{2}{*}{$200$} & $1366_{(6)(3)}^{(43)(19)}-i57_{(57)(57)}^{(42)(23)}$ & $3.7(7)(4)$ & $1.8(8)(6)$ & $2.1(5)(3)$ & $1.2(3)(3)$ & $0.8(2)(2)$\\
                        ~ & ~ & $1450(15)(11)-i15(9)(5)$ & $1.5(7)(4)$ & $3.1(5)(4)$ & $0.5(2)(1)$ & $2.1(3)(6)$ & $0.5(1)(3)$\\\cline{3-8}
    \multirow{2}{*}{2 channels}&\multirow{2}{*}{$200$} & $1387(7)(6)$ & $3.4(10)(10)$ & $1.8(5)(8)$ & $1.9(10)(5)$ & \dots & \dots \\
                        ~ & ~ & $1455(14)(6)-i29(11)(7)$ & $2.2(7)(3)$ & $3.8(6)(3)$ & $0.6(1)(1)$ & \dots & \dots 

    \\\hline
    \end{tabular}
    \caption{\label{tab:strategy1-up-to-NLO-fitting-result-st1} Pole positions and couplings of the $\Lambda(1405)$ for $m_\pi=138$ and $200$ MeV.}
\end{table*}
\begin{table*}[!tbph]
    \renewcommand{\arraystretch}{1.6}
 \setlength{\tabcolsep}{0.28cm}
    \centering
    \begin{tabular}{cccccccccc}
    \hline 
    ~ & \multicolumn{3}{c}{WT} & \multicolumn{3}{c}{LO [WT$+$Born]} & \multicolumn{3}{c}{LO$+$NLO} \\
    \cline{2-10} 
     & $z_1$ & $z_2$ & $z_3$ & $z_1$ & $z_2$ & $z_3$ & $z_1$ & $z_2$ & $z_3$ \\\cline{2-10} 

    Pole (MeV) & $1556(2)$ & $1606(1)$ & $1606(1)$ & $1548(3)$ & $1601(1)$ & $1607(1)$ & $1573(6)(6)$ & $1589(7)(5)$ & $1603(9)(10)$ \\

    $\abs{g_{(1)}}$ & $3.0(1)$ & $0$ & $0$ & $3.0(1)$ & $0$ & $0$ & $2.8(1)(1)$ & $0$ & $0$\\

    $\abs{g_{(8_a)}}$ & $0$ & $0.8(1)$ & $0$ & $0$ & $1.8(1)$ & $0$ & $0$ & $2.4(8)(2)$ & $0$ \\

    $\abs{g_{(8_b)}}$ & $0$ & $0$ & $0.8(1)$ & $0$ & $0$ & $0.4(1)$ & $0$ & $0$ & $1.7(6)(5)$ \\

  \hline 
    \end{tabular}
    \caption{\label{tab:pole-coupling-su3}Pole positions and couplings to the relevant multiplets in the SU(3) limit.}
\end{table*}

With this, we can also make a prediction for the evolution of the two poles over the $m_s=m_{s,\mathrm{phy}}$ trajectory. This is shown in Fig.~\ref{fig:trajectory}~(b). In this case, we obtain that the symmetric point corresponds to $m_\pi=758$~MeV. However, at this high pion mass, the NLO ChPT expansion breaks~\cite{Gasser:1983yg,Gasser1984}, and in particular, the expressions of the decay constants $f_\phi$, $\phi=\pi,K,\eta$ fail. Still, we can make predictions up to $450$~MeV, were the NLO ChPT relations for the masses and pseudoscalar decay constants work well~\cite{Molina2020}. These values are also consistent with previous analysis of the baryon masses~\cite{Ren2012}. Since now the $\bar{K}$ and the $\Sigma$ masses increase with the pion mass, see blue lines in Fig.~\ref{fig:baryon-mass1}, we obtain that the real parts of the two poles, which couple significantly to $\bar{K}N$ and $\pi\Sigma$, rise much faster than in the $\mathrm{Tr}[M]=C$ trajectory. In the $m_s=m_{s,\mathrm{phy}}$ trajectory, the lower pole evolves from a resonance into a virtual state at $m_\pi= 200~(178)$~MeV at LO (LO+NLO). Then, it becomes a bound state at $m_\pi=236~(278)$ MeV. The poles corresponding to the $\Lambda(1405)$ and $\Lambda(1670)$ are always found in the second Riemann sheet as resonances below $m_\pi = 450$~MeV.  These results can be tested in future LQCD simulations.

We can also compare our findings with previous works. In Ref.~\cite{Guo2023}, the energies of the two poles at the SU(3) symmetric point at LO become $1704$ and $1788$~MeV for the singlet and two octets, respectively. These values differ more than $150$~MeV to the ones obtained here at this order, while we obtain similar mass splitting between the poles coupling to the octet representations at NLO, which is $\Delta E^{(8)}=14$~MeV. However, note that the pion mass used in Ref.~\cite{Guo2023} at the SU(3) symmetric point is $512$~MeV, which differs almost $100$~MeV from the one used here in this limit. 
In addition, while in Ref.~\cite{Guo2023} is predicted that at NLO, the trajectories of the $\Lambda(1405)$ two-pole exchange with respect to the LO, we do not find here such phenomena. Our results for LO and NLO are qualitatively similar, except for the width of the lower pole which is larger at the higher order for pion masses close to the physical point. This is consistent with the dominance of the LO contribution observed in previous works~\cite{Oset1997,Jido2003,Bruns:2021krp,Lu2022}. We would like to stress that our analysis is based on the recent LQCD data on baryon masses, pseudoscalar meson masses, and decay constants along several chiral trajectories, $\mathrm{Tr}[M]=C$, $m_s=m_{s,\mathrm{phy}}$, and $m_{u(d)}=m_s$ for the CLS ensembles, while in Ref.~\cite{Guo2023}, the $b$'s are fixed to old data from Ref.~\cite{Hall2014} for only one trajectory, $m_s=m_{s,\mathrm{phy}}$, and similarly for the meson parameters~\cite{Nebreda2010}. Moreover, the scheme used for the subtraction constants is also different. While these parameters change with the scale $\mu=m_{Bi}$ in Ref.~\cite{Guo2023} (see also Refs.~\cite{Lutz2001,Hyodo2008}), we fix the scale $\mu=630$ MeV, and the subtraction constant quark mass dependence is implemented as in Ref.~\cite{Oller:2019opk}, by comparing the loop functions obtained in the cutoff and DR schemes. In addition, the values of the subtraction constants obtained in this work are close to the natural value, $a\simeq -2$~\cite{Ikeda2011}. Concretely, one gets from Eq.~(\ref{eq:subtraction-cutoff}),  $a_{\pi\Sigma}=-2.15(-2.18)$, and $a_{\bar{K}N}=-1.92(-1.89)$ for $m_\pi=200$~MeV ($138$~MeV). The subtraction constant in the SU(3) limit is $a_\mathrm{SU(3)}= -2.08$.

Indeed, the pole trajectories in the chiral unitary framework were studied first in Refs.~\cite{Jido2003,Garcia-Recio:2003ejq} at LO and in Ref.~\cite{Bruns:2021krp} up to NLO. The results of Refs.~\cite{Jido2003,Garcia-Recio:2003ejq,Bruns:2021krp} show that these trajectories at LO and up to NLO are similar. Here, we are also consistent with the qualitative trends obtained in Refs.~\cite{Jido2003,Garcia-Recio:2003ejq,Bruns:2021krp} where a simple quadratic formula for the meson and baryon masses is assumed~\footnote{In Refs.~\cite{Jido2003,Garcia-Recio:2003ejq,Bruns:2021krp}, the authors take formulas of the type, $m_i^2=m^2_0+x(m_i^2-m_0^2)$, and $M_i(x) = M_0 + x(M_i - M_0)$, where $x=0$ corresponds to the SU(3) symmetric point and $x=1$ to the physical point, $m_i$ and $M_i$ stand for the pseudoscalar meson and baryon masses in the physical point, and the subindex $0$ refers to the symmetric point.}. However, the values obtained here for the poles at the symmetric point differ significantly from the ones in Refs.~\cite{Jido2003,Garcia-Recio:2003ejq,Bruns:2021krp}. The reason is that we have used a more realistic description of the baryon masses based on the covariant chiral perturbation theory and recent LQCD data. Our work is the first one that shows agreement between both experimental and LQCD data consistently. As a result, we have made realistic predictions concerning the quark mass dependence of the two-poles of the $\Lambda(1405)$. 

Finally, we show our prediction in the $I=1$ sector below.
\subsection{Prediction in the $I=1$ sector}\label{sec:is1}
\begin{table}
 \renewcommand{\arraystretch}{1.45}
 \setlength{\tabcolsep}{0.15cm}
 \caption{\label{tab:pole-I1-coupling}Pole positions and couplings of the two $I=1$ states at the physical point.}
    \centering
    \begin{tabular}{cccccc}
    \hline
        Pole (MeV) & $|g_{\pi\Lambda}|$ & $|g_{\pi\Sigma}|$ & $|g_{\bar{K}N}|$ & $|g_{\eta\Sigma}|$ & $|g_{K\Xi}|$\\
        \hline
        $1345_{(15)}^{(63)} - i315_{(18)}^{(16)}$ & $0.3(1)$ & $2.2(2)$ & $2.3(2)$ & $0.8(4)$ & $1.6(1)$\\
        $1447_{(47)}^{(25)} - i4_{(4)}^{(20)}$ & $0.9(1)$ & $1.2(2)$ & $2.5(2)$ & $1.6(4)$ & $0.3(1)$\\
    \hline
    \end{tabular}
\end{table}
In the previous section we obtained cross sections which agree well with the  experimental ones. Then, it is reasonable to make predictions for $I=1$. In this sector, the channels contributing are, $\pi\Lambda$, $\pi\Sigma$, $\bar{K}N$, $\eta\Sigma$, and $K\Xi$. We find that there are two poles in $I=1$ located at $1345_{(15)}^{(63)}-i315_{(18)}^{(16)}$~MeV and $1447_{(47)}^{(25)}-i4_{(4)}^{(20)}$~MeV on the sheet $[--+++]$. The pole positions and couplings obtained at the physical point are shown in Table~\ref{tab:pole-I1-coupling}. These two poles couple also strongly to $\pi\Sigma$ and $\bar{K}N$ but the other channels are non-negligible. We would like to make here a remark. If only the WT term is considered, we do not find poles, but just a cusp around the $\bar{K}N$ threshold. This is in agreement with previous works based on the WT term~\cite{Jido2003,Roca2013}. 
The poles in Table~\ref{tab:pole-I1-coupling} are found when the NLO interaction is also accounted for.
In the SU(3) limit, over the $\mathrm{Tr}[M]=C$ trajectory, we find two bound states, which are located at $1589(8)$~MeV and $1603(9)$~MeV. The couplings of those to the $8_a$, and $8_b$ representations in this limit are $[2.4(9), 0]$ and $[0, 1.7(8)]$, respectively. Thus, in the SU(3) symmetric point these poles couple to the octets. The angle found here is $-\,60^\circ$. We have also evaluated the quark mass dependence of the poles. Their chiral trajectories over the $\mathrm{Tr}[M]=C$ line can be found in the Supplemental material. The interaction gets more attractive when the pion mass increases and the poles become bound in the $\pi\Sigma$ channel at $m_\pi=330$~MeV and $415$~MeV for the lower and higher energy poles, respectively. At the physical point, while the first pole is too broad, see Table~\ref{tab:pole-I1-coupling}, the second pole shows as a narrow structure close to become a cusp around the $\bar{K}N$ threshold, see the Supplemental material. This structure could be seen in the real axis and can be interpreted as a $\Sigma^*\qty(1/2^-)$ state. We find agreement with previous studies on the location of the narrow pole~\cite{Lu2022,Oller2000,Oller:2006jw,PhysRevC.100.015208,Guo2012,PhysRevD.110.114018,Li:2024rqb,Wang:2024jyk,Zhang:2004xt}. These results are interesting can be tested in future LQCD simulations.

\section{Conclusion and outlook}~\label{sec:con}
In this work, we have analyzed LQCD data on $\pi\Sigma-\bar{K}N$ scattering for $I=0$ (energy levels) at $m_\pi\simeq 200$~MeV and LQCD octect-baryon masses data in the range of pion masses between the physical point and $m_\pi=450$~MeV. The theoretical framework is based on the NLO chiral Lagrangians for the meson-baryon interaction and covariant baryon chiral perturbation theory for the baryon masses. We have also implemented the subtraction constants in the meson-baryon loops following the scheme of Ref.~\cite{Oller:2019opk}. Our analysis took into account recent LQCD data on the baryon masses, meson masses, and pseudoscalar decay constants, and the subtraction constants turn out to be of natural size in the range of pion masses considered. Therefore, we have obtained the most precise determination of the trajectories of the two-pole $\Lambda(1405)$ toward the symmetric line on the $\mathrm{Tr}[M]=C$ curve. The extrapolation of our results to the physical point is consistent with the experiment. Remarkably, our results also agree with the cross-section data. Thus, consistency between the chiral unitary approach predictions for the two-pole structure, the recent LQCD scattering data~\cite{Bulava2023,Bulava2023b}, and the experimental data, is shown for the first time. We found that both poles lie on the physical Riemann sheet at the symmetric point at $m_\pi=423$~MeV. Concretely, at the SU(3) limit, we find one of the $\Lambda(1405)$ poles coupling to the singlet and located at $E^{(1)}=1573(6)(6)$ MeV, while the higher is at $E^{(8a)}=1589(7)(5)$ MeV and couples to the octet representation. There is a third pole found connected to the $\Lambda(1670)$, shifted by $14$~MeV and belonging to the $8b$ representation. From these results, we obtain the trend over the $m_s=m_{s,\mathrm{phy}}$ curve till $m_\pi=450$~MeV, where the lower-energy pole becomes a bound state at $m_\pi=278$~MeV. Finally, we have also made predictions in the $I=1$ sector, where we find a narrow structure just below the $\bar{K}N$ threshold compatible with previous experimental analysis.

Since the pole trajectories of the scattering amplitude are connected to the nature of dynamically generated resonances through chiral dynamics and the  SU(3) breaking pattern, we believe that the results obtained here strongly support the two-pole structure of the $\Lambda(1405)$ generated from the $\pi\Sigma-\bar{K}N$ interaction. We have shown that the additional LQCD data also supports the two-pole structure consistently with the experiment. Therefore, this work significantly advances the comprehension of this long-standing puzzle, which is also essential to understanding the interaction between nuclei and strange mesons relevant to nuclear physics and astrophysics. This study can be extended to other states, such as the controversial $N^{*}(1535)$. The present work can also be tested in future LQCD simulations and experimental measurements~\cite{He:2024uau}.

%

\begin{acknowledgments} 
We are grateful to J. A. Oller, P. C. Bruns, A. Ciepl\'y, J. R. Pelaez, S. Cruz-Alzaga, and F. Gil-Dominguez for useful discussions. We acknowedge to J. Bulava and the BaSc Collaboration, and the RQCD Collaboration for making the data available to us. The plots were made with {\tt Matplotlib}~\cite{Hunter:2007} and the calculations were performed by {\tt Julia}~\cite{bezanson2017julia}. R. M. acknowledges support from the ESGENT program with Ref. ESGENT/018/2024 and the PROMETEU program with Ref. CIPROM/2023/59, of the Generalitat Valenciana, and also from 
the Spanish Ministerio de Economia y Competitividad and European Union (NextGenerationEU/PRTR) by the grant with Ref. CNS2022-13614. L. S. G. acknowledges support from the National Key R\&D Program of China under Grant No. 2023YFA1606700. This work is also partly supported by the Spanish Ministerio de Economia y Competitividad (MINECO) and
European FEDER funds under Contracts No. FIS2017-84038-C2-1-P B, PID2020-112777GB-I00, and by Generalitat Valenciana under contract PROMETEO/2020/023. This project has received funding from the European Union Horizon 2020 research
and innovation program under the program H2020-INFRAIA-2018-1, grant agreement No. 824093 of the STRONG-2020
project.
\end{acknowledgments}

%


\clearpage
\begin{widetext}
\section*{Supplemental material}
This supplemental material presents details regarding the fitting procedure and several intermediate results, namely, the inelasticities and phase shifts, the correlation matrix of the fit to the energy levels, the fit to the octet baryon masses, the fitting results at LO, and the predicted trajectories of the poles in $I=1$.
\subsection{Phase shifts and inelasticities}
Inelasticities $\eta$ and phase shifts $\delta_{\pi\Sigma}$ and $\delta_{\bar{K}N}$ up to NLO as a function of the center-of-mass energy difference with respect to the $\pi\Sigma$ threshold is given in Fig.~\ref{fig:phase-shift-up-to-nlo-st1} for the two-channel case with WT term (left), two and four channel up to NLO (center and right panels respectively). The phase shifts agree well with the findings in the LQCD simulation~\cite{Bulava2023,Bulava2023b}. The $\pi\Sigma$ phase shift rise more smoothly in the four-coupled channel calculation analysis performed here. When the $\eta\Lambda$ and $K\Xi$ channels are removed, an small bump close to the $\pi\Sigma$ threshold is produced, more pronounced when only the WT term is kept. This bump is related to the fact that a virtual bound state is found in these cases. The inelasticity increases for higher energies when the WT term is included, similarly to the LQCD simulation~\cite{Bulava2023,Bulava2023b}. 
\begin{figure}[!htbp]
\centering
    \includegraphics[width=0.95\textwidth]{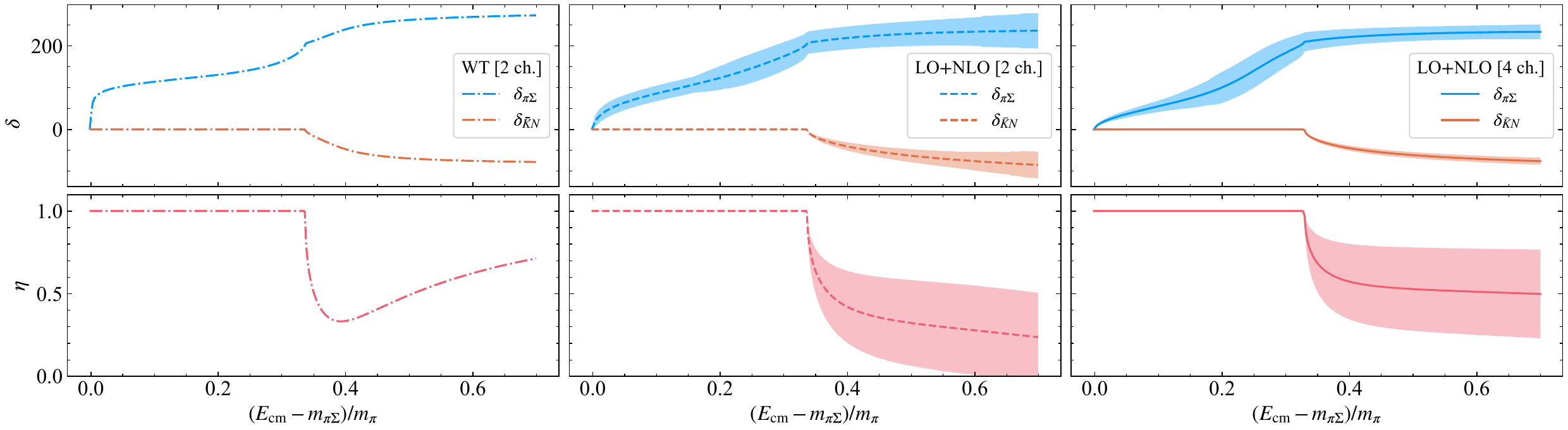}
\caption{\label{fig:phase-shift-up-to-nlo-st1}Inelasticities$\eta$ and phase shifts $\delta_{\pi\Sigma}$ and $\delta_{\bar{K}N}$ up to NLO as a function of the center-of-mass energy difference with respect to the $\pi\Sigma$ threshold.}
\end{figure}
\subsection{The correlation matrix of the fit to the energy levels}
The heat map of the correlation matrix between the LECs is given in Fig.~\ref{fig:heatmap-LECs-qmax}.
\begin{figure}[!htpb]
	\includegraphics[width=0.4\textwidth]{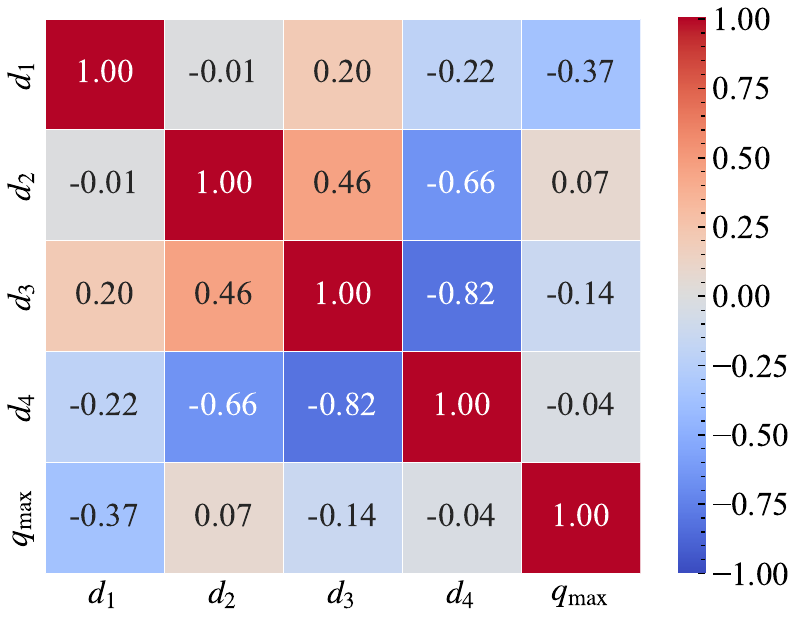}
	\caption{\label{fig:heatmap-LECs-qmax}Heat map of the correlation matrix for the LECs from the fit to the energy levels.}
\end{figure}







\subsection{The LECs of of the interaction kernels in the irreducible representations}
\subsubsection{$I=0$}
We project the isospin bases of $I=0$ to the irreducible representations $\{1, 8, 8', 27\}$, i.e.,
\begin{equation}
\renewcommand{\arraystretch}{2} 
	\begin{bmatrix}
		\ket{\pi\Sigma} \\%
		\ket{\bar{K}N} \\%
		\ket{\eta\Lambda}\\%
		\ket{K\Xi}
	\end{bmatrix}
	=
	\begin{bmatrix}
		\frac{\sqrt{\frac{3}{2}}}{2} & -\sqrt{\frac{3}{5}} & 0 & -\frac{1}{2 \sqrt{10}} \\
 -\frac{1}{2} & -\frac{1}{\sqrt{10}} & \frac{1}{\sqrt{2}} & -\frac{\sqrt{\frac{3}{5}}}{2} \\
 -\frac{1}{2 \sqrt{2}} & -\frac{1}{\sqrt{5}} & 0 & \frac{3 \sqrt{\frac{3}{10}}}{2} \\
 \frac{1}{2} & \frac{1}{\sqrt{10}} & \frac{1}{\sqrt{2}} & \frac{\sqrt{\frac{3}{5}}}{2} \\
	\end{bmatrix}
	\begin{bmatrix}
		\ket{1} \\
		\ket{8} \\
		\ket{8'} \\
		\ket{27}
	\end{bmatrix}.
\end{equation}
In the SU(3) limit, the LECs of the interaction kernels for the $I=0$ up to NLO are given by
\begin{align}
    \renewcommand{\arraystretch}{2} 
	C^{\mathrm{WT}}_{\mathrm{SU(3)}} &= 
    \begin{bmatrix}
        6 & ~ & ~ & ~ \\
        ~ & 3 & ~ & ~ \\
        ~ & ~ & 3 & ~ \\
        ~ & ~ & ~ & -2
    \end{bmatrix},\\[10pt]
    C^{\mathrm{Born-}s}_{\mathrm{SU(3)}} &= 
    \begin{bmatrix}
        0 & 0 & 0 & 0 \\
        0 & \frac{10D^2}{3} & -2\sqrt{5}DF & 0 \\
        0 & -2\sqrt{5}DF & 3F^2 & 0 \\
        0 & 0 & 0 & 0
    \end{bmatrix},\,
    C^{\mathrm{Born-}u}_\mathrm{SU(3)} = 
    \begin{bmatrix}
        \frac{10D^2}{3}-6F^2 & 0 & 0 & 0 \\
        0 & -D^2-3F^2 & 0 & 0 \\
        0 & 0 & 3F^2-\frac{5D^2}{3} & 0 \\
        0 & 0 & 0 & \frac{2}{3}\qty(D^2 + 3F^2)
    \end{bmatrix}, \\[10pt]
    C^{\mathrm{NLO1}}_\mathrm{SU(3)} & = 
    \begin{bmatrix}
        \frac{4}{3}m^2\qty(3b_0 + 7b_D) & 0 & 0 & 0 \\
        0 & \frac{2}{3}m^2\qty(6b_0 + b_D) & -2\sqrt{5}m^2b_F & 0 \\
        0 & -2\sqrt{5}m^2b_F & 2m^2\qty(2b_0 + 3b_D) & 0 \\
        0 & 0 & 0 & 4m^2\qty(b_0 + b_D)
    \end{bmatrix},\\[10pt]
    C^\mathrm{NLO2}_\mathrm{SU(3)} &= 
    \begin{bmatrix}
        -6d_2+9d_3+2d_4 & 0 & 0 & 0 \\
        0 & -3d_2+d_3+2d_4 & -\sqrt{5}d_1 & 0 \\
        0 & -\sqrt{5}d_1 & 9d_2-d_3+2d_4 & 0 \\
        0 & 0 & 0 & 2d_2 + d_3 + 2d_4 
    \end{bmatrix},
\end{align}
where $m$ is the mass of the pseudoscalar meson in the SU(3) limit.
\subsubsection{$I = 1$}
The isospin $I=1$ bases in the irreducible representations $\{8, 8', 10, \bar{10}, 27\}$ are given by
\begin{equation}
    \renewcommand{\arraystretch}{2} 
    \begin{bmatrix}
        \ket{\pi\Lambda} \\
        \ket{\pi\Sigma} \\
        \ket{\bar{K}N} \\
        \ket{\eta\Sigma} \\
        \ket{K\Xi}
    \end{bmatrix}
    =
    \begin{bmatrix}
        \frac{1}{\sqrt{5}} & 0 & \frac{1}{2} & \frac{1}{2} & \sqrt{\frac{3}{10}} \\
        0 & \sqrt{\frac{2}{3}} & \frac{1}{\sqrt{6}} & -\frac{1}{\sqrt{6}} & 0 \\
        -\sqrt{\frac{3}{10}} & -\frac{1}{\sqrt{6}} & \frac{1}{\sqrt{6}} & -\frac{1}{\sqrt{6}} & \frac{1}{\sqrt{5}} \\
        \frac{1}{\sqrt{5}} & 0 & -\frac{1}{2} & -\frac{1}{2} & \sqrt{\frac{3}{10}} \\
        -\sqrt{\frac{3}{10}} & \frac{1}{\sqrt{6}} & -\frac{1}{\sqrt{6}} & \frac{1}{\sqrt{6}} & \frac{1}{\sqrt{5}} \\
    \end{bmatrix}
    \begin{bmatrix}
        \ket{8} \\
        \ket{8'} \\
        \ket{10} \\
        \ket{\bar{10}} \\
        \ket{27}
    \end{bmatrix}.
\end{equation}
The LECs of $I=1$ in the irreducible representations up to NLO are given by
\begin{align}
    \renewcommand{\arraystretch}{2} 
    C_\mathrm{SU(3)}^\mathrm{WT} &=
    \begin{bmatrix}
        3 & 0 & 0 & 0 & 0 \\
        0 & 3 & 0 & 0 & 0 \\
        0 & 0 & 0 & 0 & 0 \\
        0 & 0 & 0 & 0 & 0 \\
        0 & 0 & 0 & 0 & -2
    \end{bmatrix},
    \\[10pt]
    C_\mathrm{SU(3)}^{\mathrm{Born-}s} &=
    \begin{bmatrix}
        \frac{10}{3}D^2 & -2\sqrt{5}DF & 0 & 0 & 0 \\
        -2\sqrt{5}DF & 6F^2 & 0 & 0 & 0 \\
        0 & 0 & 0 & 0 & 0\\
        0 & 0 & 0 & 0 & 0\\
        0 & 0 & 0 & 0 & 0
    \end{bmatrix},\,
    C_\mathrm{SU(3)}^{\mathrm{Born-}u}=
    \begin{bmatrix}
        -D^2-3F^2 & 0 & 0 & 0 & 0\\
        0 & 3F^2-\frac{5D^2}{3} & 0 & 0 & 0 \\
        0 & 0 & \frac{4}{3}D\qty(D+3F) & 0 & 0 \\
        0 & 0 & 0 &\frac{4}{3}D\qty(D - 3F) & 0 \\
        0 & 0 & 0 & 0 & \frac{2}{3}\qty(D^2 + 3F^2)
    \end{bmatrix},\\[10pt]
    C_\mathrm{SU(3)}^\mathrm{NLO1} &= 
    \begin{bmatrix}
        \frac{2}{3}m^2\qty(6b_0+b_D) & -2\sqrt{5}m^2b_F & 0 & 0 & 0 \\
        -2\sqrt{5}m^2b_F & 2m^2\qty(2b_0+3b_D) & 0 & 0 & 0 \\
        0 & 0 & 4m^2\qty(b_0+b_F) & 0 & 0 \\
        0 & 0 & 0 & 4m^2\qty(b_0-b_F) & 0 \\
        0 & 0 & 0 & 0 & 4m^2\qty(b_0+b_D)
    \end{bmatrix},\\[10pt]
    C_\mathrm{SU(3)}^\mathrm{NLO2} &=
    \begin{bmatrix}
        -3d_2+d_3+2d_4 & -\sqrt{5}d_1 & 0 & 0 & 0 \\
        -\sqrt{5}d_1 & 9d_2-d_3+2d_4 & 0 & 0 & 0 \\
        0 & 0 & 2d_1-d_3+2d_4 & 0 & 0 \\
        0 & 0 & 0 & -2d_1-d_3+2d_4 & 0 \\
        0 & 0 & 0 & 0 & 2d_2+d_3+2d_4
    \end{bmatrix}.
\end{align}
\subsection{Meson masses\label{sec:pseu_meson}}
The pseudoscalar meson masses up to NLO in ChPT are given by~\cite{Gasser1984}
\begin{equation}
    m_\pi^2 = M_{0\,\pi}^2\left[1+\mu_\pi-\frac{\mu_\eta}{3}+\frac{16 M_{0\,K}^2}{f_0^2}\left(2L_6^r-L_4^r\right)\right]+\frac{8 M_{0\,\pi}^2}{f_0^2}\left(2L_6^r+2L_8^r-L_4^r-L_5^r\right)\,, \label{eq:pimass}
\end{equation}

\begin{equation}
    m^2_K = M^2_{0\,K}\left[1+\frac{2\mu_\eta}{3}+\frac{8 M_{0\,\pi}^2}{f_0^2}\left(2L_6^r-L_4^r\right)+\frac{8 M_{0\,K}^2}{f_0^2}\left(4L_6^r+2L_8^r-2L_4^r-L_5^r\right)\right]\,, \label{eq:kmass}
\end{equation}
\begin{equation}
    \begin{aligned}
        m^2_\eta = & M^2_{0\,\eta} \left[1+2\mu_K-\frac{4}{3}\mu_\eta+\frac{8M^2_{0\,\eta}}{f_0^2}(2L_8^r-L_5^r)
        +\frac{8}{f_0^2}(2 M^2_{0\,K}+M^2_{0\,\pi})(2L_6^r-L_4^r)\right] \\
        &+ M^2_{0\,\pi}\left[-\mu_\pi+\frac{2}{3}\mu_K+\frac{1}{3}\mu_\eta\right]+\frac{128}{9f_0^2}(M^2_{0\,K}-M^2_{0\,\pi})^2(3L_7+L_8^r)\,,\label{eq:etamass}
    \end{aligned}
    \end{equation}
    with
    \begin{equation}\label{eq:tadpole}
    \mu_P=\frac{M_{0\, P}^2}{32 \pi^2 f_0^2}\log\frac{M_{0 \,P}^2}{\mu^2_r},\qquad P=\pi,K,\eta\,.
    \end{equation}
    The superscript $r$ denotes renormalized LECs, which carry the dependence on the regularization scale $\mu_r$~\cite{Gasser1984}. In the above equation, $M_{0P}$, with $P=\pi,K,\eta$, represent the pseudoscalar meson masses at LO and $f_0$ the decay constant in the chiral limit. Here, we take $\mu_r=770$ MeV and $f_0=80$ MeV as in Ref.~\cite{Molina2020}. The LECs $L_i$ are taken from Table X of~\cite{Molina2020}.

\subsection{Baryon masses\label{sec:baryon-mass-lattice}}
We analyze the data from Ref.~\cite{RQCD2022} for the baryon masses of the CLS ensembles using the one-loop NLO covariant baryon chiral perturbation theory~\cite{Jorge2010}. 
\begin{equation}\label{eq:octet-baryon-mass}
    m_B = m_0 + m_B^{(2)} + m_B^{(3)}\ ,
\end{equation}
where
\begin{equation}\label{eq:LO-baryon-mass}
    m_B^{(2)} = \sum_{\phi=\pi,K} -\xi_{B,\phi}^{(a)} m_\phi^2,
\end{equation}
and
\begin{equation}\label{eq:one-loop-baryon-mass}
    m_B^{(3)} = \sum_{\phi=\pi,K,\eta}\frac{1}{(4\pi f_\phi)^2}\xi_{B,\phi}^{(b)} H_B^{(b)}(m_\phi)\ ,
\end{equation}
where $m_0$ stands for the baryon mass in the chiral limit, and $m_B^{(2)}$ and $m_B^{(3)}$ represent the polynomial and one-loop contributions, respectively. The coefficients $\xi_{B,\phi}^{(a)}$ and $\xi_{B,\phi}^{(b)}$ can be found in Ref.~\cite{Jorge2010}. These are combinations of the pertinent LECs, $b_0, b_D, b_F, D,F$. The EOMS loop-function $H_B^{(b)}(m_\phi)$ can be found in Ref.~\cite{RQCD2022}. The baryon axial coupling constants $D$ and $F$ are fixed to be~\cite{Borasoy1998},
\begin{equation}\label{eq:axial-coupling-constant}
    D=0.80, \ F=0.46.
\end{equation}
In Ref.~\cite{RQCD2022} there are three types of chiral trajectories, i.e., $\mathrm{Tr}\qty[M]=C$, $m_s=m_{s,\mathrm{phy}}$, and $m_s=m_l$, where $l=u,\,d$. In Eqs.~\eqref{eq:pimass}-\eqref{eq:kmass}, the pseudoscalar mesons' masses depend on $m_{0\pi}$ and $m_{0K}$. The Gell-Mann-Okubo relation gives
\begin{equation}
	m_{0K}^2 = B_0C - \frac{m_{0\pi}^2}{2},\,
	m_{0K}^2 = B_0m_{s,\mathrm{phy}} + \frac{m_{0\pi}^2}{2}
\end{equation}
with respect to $\mathrm{Tr}[M]=C$ and $m_s=m_{s,\mathrm{phy}}$, respectively. The free parameters needed to fit the hadron masses in Ref.~\cite{RQCD2022} are $B_0C$, $B_0m_s$, $m_0$, $b_0$, $b_D$, and $b_F$. The results of the fits are plotted in Fig.~\ref{fig:baryon-mass}. In Ref.~\cite{RQCD2022}, the authors have corrected the systematic effects, e.g.,  the finite volume and finite lattice spacing. The details are explained in Sec. 4.2 of Ref.~\cite{RQCD2022}. We have taken the data from Ref.~\cite{RQCD2022} already corrected for the systematic effects, and these are the data that we have used. The covariance matrix for the octet baryon masses and the pseudo-scalar meson masses is also included in this fit. Indeed, in Ref.~\cite{RQCD2022}, the authors also perform a fit of the baryon masses using the Baryon Chiral Perturbation theory up to $\mathcal{O}(p^3)$. The reason why the LECs from Ref.~\cite{RQCD2022} are not used here is because the parameters $D$ and $F$ in Eq. (18) of Ref.~\cite{RQCD2022} are also fitted to the data. However, the constants $D$ and $F$ should be determined by fitting to the semi-leptonic decays $B\to B'+e^- + \bar{\nu}_e$ at tree level~\cite{Borasoy1998}. For this reason, we have fixed $D$ and $F$ to the values given in Ref.~\cite{Borasoy1998}. The values that we obtain from the fit are given in Table~\ref{tab:LECs-octet-baryon}, where we also show the LECs obtained in Ref.~\cite{RQCD2022} for comparison. The correlations between the parameters are given in Fig.~\ref{fig:heatmap-parameter-hadron-mass}. In particular, we observe a very high correlation between the $m_0$ and $b_0$ parameters. It can be seen that the LECs obtained here are compatible within uncertainties with the ones given in Ref.~\cite{RQCD2022}.

\begin{table}[!ht]
\renewcommand{\arraystretch}{1.6}
 \setlength{\tabcolsep}{0.26cm}
\caption{\label{tab:LECs-octet-baryon}Free parameters fixed by the hadron masses in Ref.~\cite{RQCD2022}. The first uncertainty is statistical, and the second comes from the uncertainty in the lattice spacing~\cite{RQCD2022}. Our results are compared with those obtained in Ref.~\cite{RQCD2022}.}
\begin{tabular}{lcccccc}
    \hline
    ~ & $\sqrt{B_0C}$ (MeV) & $\sqrt{B_0m_{s,\mathrm{phy}}}$ (MeV) & $m_0$ (MeV) & $b_0\,\qty(\mathrm{GeV}^{-1})$ & $b_D\,\qty(\mathrm{GeV}^{-1})$ & $b_F\,\qty(\mathrm{GeV}^{-1})$ \\
    \hline
    This work & $488(38)(42)$ & $474(50)(55)$ & $805(40)(40)$ & $-0.665(40)(28)$ & $0.062(26)(8)$ & $-0.354(18)(9)$ \\
    BChPT FV~\cite{RQCD2022} & $\cdots$ & $\cdots$ & $821_{(53)}^{(71)}$ & $-0.739_{(84)}^{(70)}$ & $0.056_{(39)}^{(43)}$ & $-0.44_{(26)}^{(40)}$ \\
    \hline
\end{tabular}
\end{table}
\begin{figure}[!htbp]
    \includegraphics[width=0.9\textwidth]{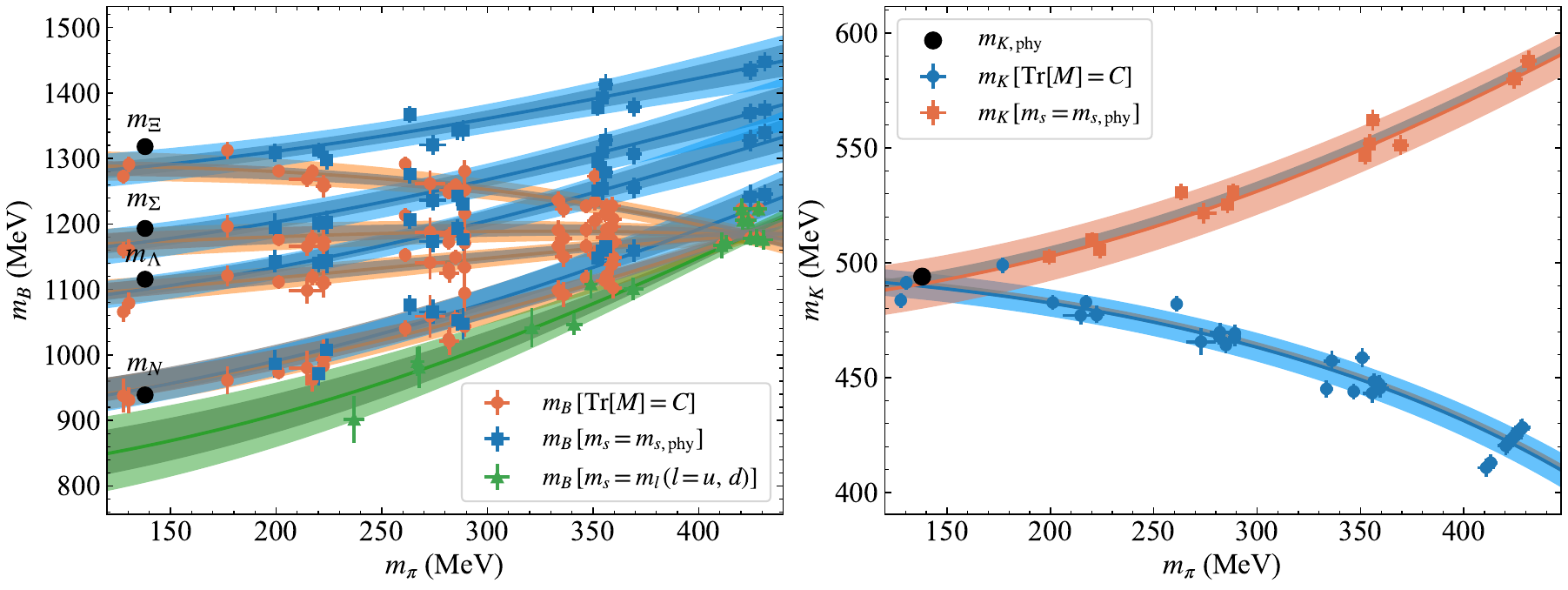}
\caption{\label{fig:baryon-mass}Fits to the hadron masses. The data is provided by RQCD~\cite{RQCD2022}. The circles, squares, and triangles denote the $\mathrm{Tr}\qty[M]=C$, $m_s=m_{s,\mathrm{phy}}$ trajectories, and the symmetric line $m_s=m_l$, where $l=u,d$, respectively. The circles in black are the physical masses of the hadrons. The colored error bands consider the systematic and statistical errors of the hadron masses. The error bands in gray are the results where only the statistical errors are considered.}
\end{figure}
\begin{figure}[!htbp]
	\includegraphics[width=0.4\textwidth]{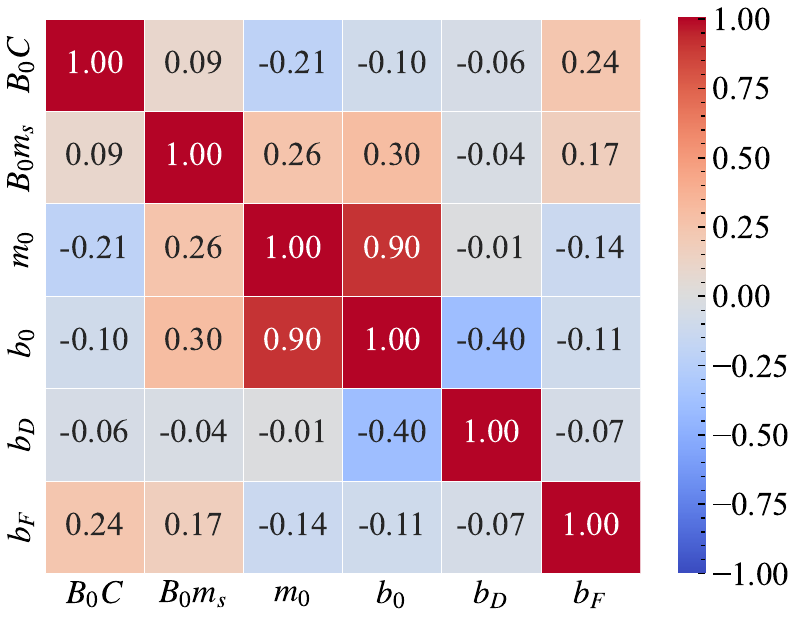}
	\caption{\label{fig:heatmap-parameter-hadron-mass}Correlations between the parameters obtained from the fit to the octet baryon masses.}
\end{figure}

\subsection{\label{sec:cutoff-LO}Results for the LO fit}
At LO, we do a fit with WT interaction and WT + Born interaction. The free parameter is the cutoff $q_\mathrm{max}$. We follow the scheme of Ref.~\cite{Oller:2019opk} for $a(\mu)$. The reduced-$\chi^2$ obtained is $3.8$ and $2.6$ for WT interaction and WT+Born interaction, respectively. The cutoff value obtained is 610(1) MeV and $669(20)$ MeV, respectively. The pole positions of the $\Lambda(1405)$ for $m_\pi=138$, and $200$ MeV are given in Table~\ref{tab:pole-position-LO-st1}.
\begin{table}[!tbph]
    \renewcommand{\arraystretch}{1.6}
 \setlength{\tabcolsep}{0.26cm}
    \centering
    \caption{\label{tab:pole-position-LO-st1}The pole positions of the $\Lambda(1405)$ for $m_\pi=138$ and $200$ MeV at LO.}
    \begin{tabular}{ccc}
    \hline 
    $m_\pi$ (MeV) & $z^{(\mathrm{WT})}$ (MeV) & $z^{(\mathrm{LO})}$ (MeV) \\\hline
    \multirow{2}{*}{$138$} & $1386(1) - i57(1)$ & $1382(2) - i41(3)$ \\
    ~ & $1437(1) - i15(1)$ & $1456(1) - i30(2)$ \\\cline{2-3}
    \multirow{2}{*}{$200$}  & $1389(1)-i21(1)$ & $1390(4)$\\
    ~ & $1471(1) - i12(1)$ & $1485(1)-i19(1)$ 
    \\\hline
    \end{tabular}
\end{table}
%




\subsection{The state with $I=1$}

    
In Fig.~\ref{fig:traj-I1-TrM}, we show the two poles predicted here in $I=1$ related to the $\Sigma^*$ in comparison with other works, Refs.~\cite{Lu2022,Oller2000,PhysRevC.100.015208,Guo2012} in the left panel, and the trajectories of the two poles over the $\mathrm{Tr}[M]=C$ curve in the right panel. The lower pole transitions from the $[--+++]$ sheet to the $[-++++]$ one at $m_\pi=330$~MeV. Then, it becomes a bound state for $m_\pi=391$~MeV. The higher pole evolves from a resonance to a bound state at $m_\pi=415$~MeV. In Fig.~\ref{fig:tmat-I1}, we show the $\abs{T_{22}}^2$, $\abs{T_{23}}^2$, and $\abs{T_{33}}^2$ matrix elements for $I=1$ which correspond to the transitions $\pi\Sigma\to\pi\Sigma, \bar{K}N$ and $\eta\Sigma\to\eta\Sigma$, respectively.
\begin{figure}[!htbp]
    \resizebox{0.4\textwidth}{!}{\import{}{./figure/poleI1_comparsion.pgf}}
    \hspace{0.4in}
    \includegraphics[width=0.4\textwidth]{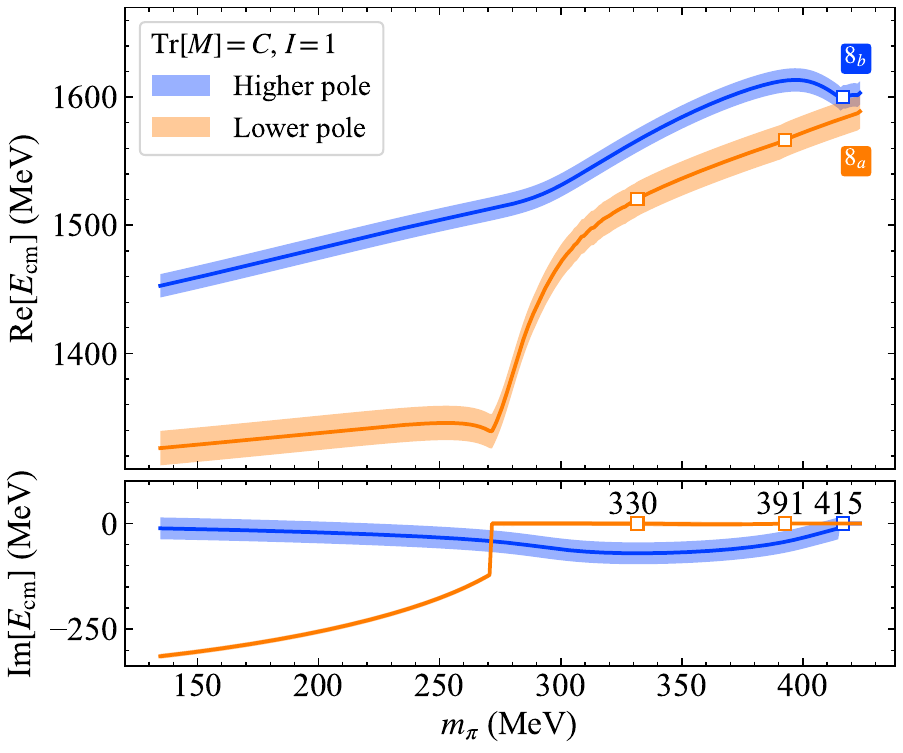}  
    \caption{\label{fig:traj-I1-TrM} The predicted poles of $I=1$ in comparsion with Refs.~\cite{Lu2022,Oller2000,PhysRevC.100.015208,Guo2012}  and the trajectories of the two poles constrained by $\mathrm{Tr}[M]=C$.}
\end{figure}
\begin{figure}
    \centering
    \includegraphics[width=0.5\linewidth]{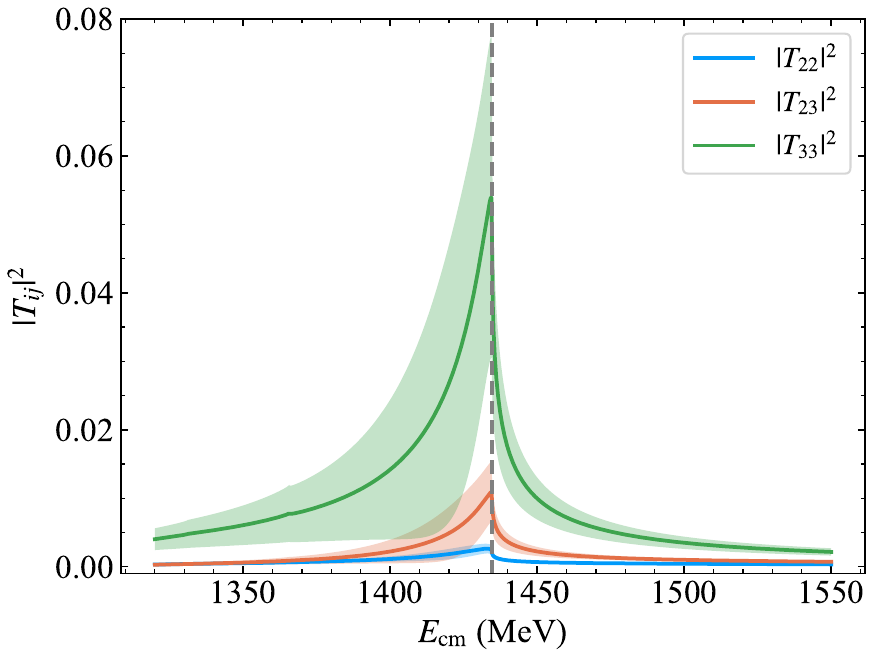}       
    \caption{Squared of the $T$-matrix elements in $I=1$ for the main contributing channels.}
    \label{fig:tmat-I1}
\end{figure}
\end{widetext}

\end{document}